\documentclass[twocolumn,showpacs,amsmath,amssymb,aps,pra,floatfix,nofootinbib,superscriptaddress]{revtex4-1}
\usepackage{graphicx,graphics,times,color}
\usepackage{bm}
\usepackage{longtable}

\def\be{\begin{equation}}
\def\ee{\end{equation}}
\def\ba{\begin{eqnarray}}
\def\ea{\end{eqnarray}}

\def\ra{\rangle}

\begin{document}

\title{Long distance entanglement generation through coherent directed transport of Neutral atoms in unmodulated optical lattices}
\author{Morteza Rafiee}
\affiliation{Department of Physics, University of Shahrood, 3619995161, Shahrood, Iran}
\author{Abolfazl Bayat}
\affiliation{Institut f\"{u}r Theoretische Physik, Albert-Einstein Allee 11, Universit\"{a}t Ulm, 89069 Ulm, Germany}
\affiliation{Department of Physics and Astronomy, University College
London, Gower St., London WC1E 6BT, United Kingdom}

\date{\today}

\begin{abstract}
We introduce a fully coherent way for directed transport of
localized atoms in optical lattices by regularly performing phase
shifts on the lattice potential during the free evolution of the
system. This paves the way for realizing a possible cold atom
quantum computer in which entangling gates operate by bringing two
individual atoms in the proximity of each other and letting them
to interact. The speed of our protocol is determined by the tunneling amplitudes of the atoms and thus is
much faster than the speed of the dynamics resulted from
superexchange interaction in spin chains. Our scheme is robust
against possible imperfections and perhaps its main advantage is
its simplicity where all of its requirements have been already
achieved in recent experiments.
\end{abstract}

\vspace*{10pt}



\maketitle

\section{Introduction} \label{sec1}
Mediating interaction between arbitrary
pairs in a network of qubits is essential for realizing two-qubit
entangling gates and thus for universal quantum computation
\cite{Bremner}.
Spin chains are the most common quantum buses where their
evolution mediate quantum gates between distant points
\cite{albanese,Yung-bose,Lukin-gate,WojcikLKGGB,bayat-gate}. However, to
avoid their dispersion one has to either delicately engineer the
couplings \cite{Yung-bose,bayat-gate} or switch to super slow
perturbative regimes \cite{Lukin-gate,WojcikLKGGB}. To have both
the flexibility and the speed one may think of using mobile qubits
as interaction mediators \cite{Cirac-bus}.
To fulfil this idea one invokes that: (i) the mobile qubit should
always remain \emph{localized} during its transport to selectively
interact with ceratin qubits; (ii) the transport itself should be
deterministic to allow for \emph{set times} at which the
interaction with the qubits in the network should be switched on.
These conditions are ultimately essential for a quantum computer
and in this perspective the main goal of this paper is to propose
a coherent mechanism for deterministic directed transport of
individual atoms in a simply realizable setup.

Directed transport of atoms in the absence of external biased
forces in a periodic structure, known as ratchet effect,
has been achieved by breaking certain temporal and spatial
symmetries of the Hamiltonian with the means AC-driving fields in both dissipative \cite{Hanggi,Renzoni-PRL} and
non-dissipative systems \cite{Salgar-science,Denisov}.
However, in these scenarios one can only talk about a net current
of an atomic cloud ($\sim 10^4$ atoms) which is quantified by the
velocity of its center of mass
\cite{Hanggi,Renzoni-PRL,Salgar-science,Denisov} and the resulted dynamics
is not ``deterministic" for individual atoms and they do not
remain ``localized" during the transport process. One may also use
AC-driving fields to renormalize
the tunneling amplitudes \cite{creffield-normalized}, without breaking the symmetries of the Hamiltonian, which then
can be used for controlled transport of localized atoms
\cite{creffield-gate,Bloch-AC-driving} or quantum states
\cite{Isart,quantum-rachet}, however, the price is a complicated setup
and a slow dynamics caused by the emergence of the weak renormalized
couplings. Optimal control theory can also be used to improve the
speed of the AC-driven transport \cite{Calarco-rachet}.

Cold atoms in optical lattices is now a very fast developing
field \cite{Bloch-Review}. Implementation of both spin s = $\frac{1}{2}$ ladders and s = $1$ chains with cold atoms in an optical lattice has been introduced \cite{Delgado}. Moreover, generating a Mott insulator phase, with exactly one atom
per site, is an ordinary experiment now \cite{Mott-insulator}
which enables for realizing effective spin Hamiltonians
\cite{Lukin} via superexchange interaction. Since, this is a
second order process the effective spin couplings are $4J^2/U$
(where, $J$ is the tunneling and $U$ is the onsite energy). These
weak couplings not only result in very slow dynamics
\cite{bayat-gate,bloch-phase-shift,Bloch-AC-driving} but also
demand very low temperatures for observing magnetic phases
\cite{Medley2010}. To avoid slow dynamics, new experiments get
benefits of the the direct tunneling of particles, which are
controlled by $J$ instead of $4J^2/U$
\cite{Bloch-AC-driving,greiner-AFM}. Moreover, single site
resolution in current experiments \cite{single-site,Weitenberg}
allows for single qubit operations and measurements
\cite{Meschede,Weitenberg,Gibbons-Nondestructive,Jaeyoon-single-site}. A series of
multiple two-qubit gates, acting globally and simultaneously,  has
been also realized \cite{Mandel-gate} between neighboring atoms.
Unfortunately, generalization of this method to distant qubits is
restricted by decoherence to only $\sim 10$ sites
\cite{Meschede-gate}. Furthermore, fast generation of a quantum register and creation of the entangled states by dynamically manipulating the shape of the lattice potential in optical superlattices has been investigated \cite{vaucher}.

In this paper we study the transport of neutral atoms in optical lattices
through iterative swapping procedure by regularly performing
phase shifts on the lattice potential. In Ref.~\cite{Isart} the swapping procedure is realized through time modulation of the intensity of the laser beams which gives essentially the same dynamics, as performing phase shifts, within a double well. Our proposal has the advantage of its simplicity for performing phase shifts. However, apart from the technical issues, there are certain points which have been remained untouched for performing iterative swapping, either realized by phase shifts or amplitude modulation, in a realistic scenario. We give a comprehensive analysis for a number of imperfections which have not yet been addressed in the literature.


The structure of this paper is as follows: in section (\ref{sec2}) we introduce our model. In section (\ref{sec3}) directed transport is explained. The time scale of our mechanism is discussed in section (\ref{sec4}) and several possible imperfections are investigated in section (\ref{sec5}). A comparison with other transport mechanisms is given in section (\ref{sec6}) and a method for entanglement detection is introduced in section (\ref{sec7}) . We finally summarize our results in section (\ref{sec8}) and acknowledgement is in section (\ref{sec9}).


\begin{figure} \centering
    \includegraphics[width=7cm,height=5cm,angle=0]{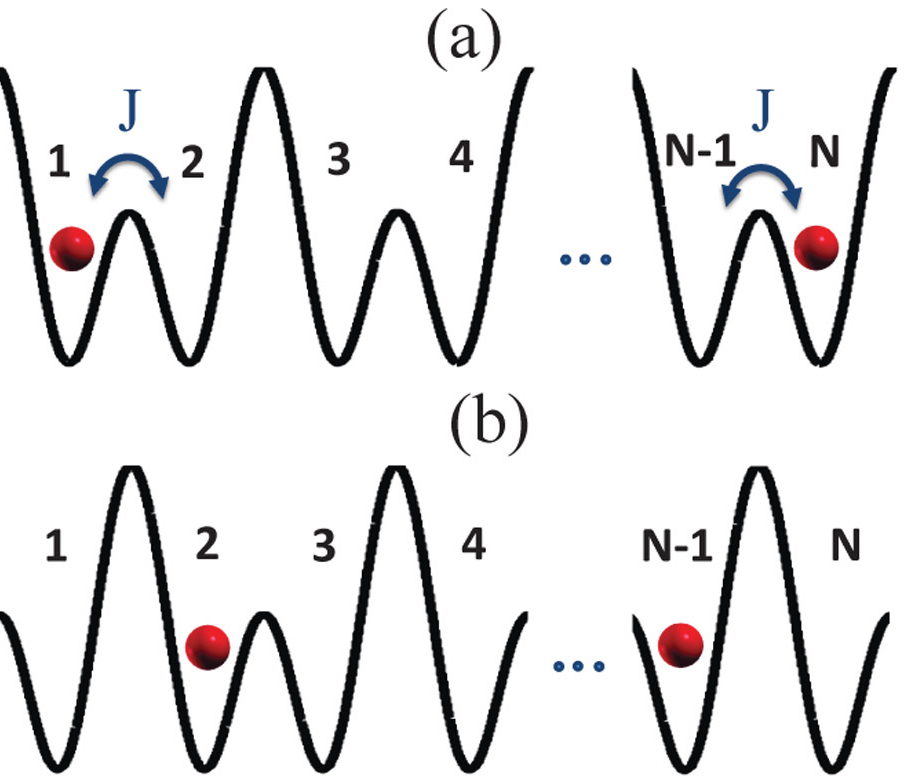}
    \caption{(Color online) (a) A superlattice containing one atom in site $1$, $N$ and the tunnelings $J$.
    (b) Applying a phase shift, after accomplishing the tunneling, changes the lattice potential and exchanges the odd and even tunnelings.}
     \label{fig1}
\end{figure}

\section{Introducing the Model} \label{sec2}

We consider an optical superlattice formed by two different set
of counter propagating laser beams where one frequency is twice
the other. The resulting potential is
\begin{equation}\label{V_lattice}
    V(x)=V_l \cos^2(2\pi x/ \lambda_l)+V_s \cos^2(2\pi x/\lambda_s+\phi)
\end{equation}
where, $\lambda_l=2\lambda_s$ are the wave lengths, $V_l$ and
$V_s$ are the amplitudes and $\phi$ is the phase shift between
laser beams. This potential has already been used in AC-driven
lattices for controlling the couplings \cite{Bloch-AC-driving}.
A superlattice can also be realized by combining different polarizations of a single frequency laser beam \cite{Sebby-Strabley}.
The low energy Hamiltonian of atoms trapped by $V(x)$ is
\cite{Lukin}
\begin{eqnarray}\label{Hubbard}
   H&=&-\sum_{<i,j>,\sigma} (J_i a^\dagger_{i,\sigma} a_{j,\sigma}+H.C.) + \frac{U}{2} \sum_{i,\sigma}
    n_{i,\sigma}(n_{i,\sigma}-1) \cr
    &+& \frac{U}{2} \sum_i n_{i,\uparrow}n_{i,\downarrow},
\end{eqnarray}
where, $<i,j>$ denotes the nearest neighbor sites, $a_{i,\sigma}$
annihilates one atom with spin $\sigma=\uparrow,\downarrow$ at
site $i$, and $n_{i,\sigma}=a^\dagger_{i,\sigma} a_{i,\sigma}$.
The tunneling $J_i$'s are controlled by the amplitudes $V_l$ and
$V_s$ such that all even (odd) couplings take the same amplitude
$J_e$ ($J_o$). Changing the phase shift $\phi$ from $0$ to $\pi/2$
displaces the lattice potential by $\lambda_s$ and thus exchanges
the even and odd couplings. We assume that the amplitudes $V_l$
and $V_s$ are tuned such that we get $J_o=J$ and $J_e=0$ for
$\phi=0$ (see Fig.~\ref{fig1}(a)) and $J_o=0$ and $J_e=J$ for
$\phi=\pi/2$ (see Fig.~\ref{fig1}(b)). This choice of couplings
make the system a series of decoupled double wells with
independent dynamics.

\section {Directed Transport} \label{sec3}

Here we consider a superlattice
potential with $\phi=0$ for which $J_o=J$ and $J_e=0$. As the
result, a single atom localized at a single site, is subjected to
a double well potential as shown in Fig.~\ref{fig1}(a).
 The dynamics of this system has been first analyzed in Ref.~\cite{Isart} for both noninteracting and interacting atoms in a double-well problem. Furthermore, optimal quantum control have been investigated for smoothly increase and decrease of the
tunneling barriers. In this section we review the results of Ref.~\cite{Isart} to determine the notation for the rest of the paper.
We denote
the state of the atom localized in the left (right) site as
$|\sigma,0\ra$ ($|0,\sigma\ra$) where, $\sigma$ represents the
spin of the atom and $0$ denotes an empty site. The evolution of
these states in a double well are given by
\begin{eqnarray}\label{single-site-evolution}
    |\sigma,0\ra &\rightarrow&  \cos(Jt) |\sigma,0\ra + i \sin(Jt) |0,\sigma\ra \cr
    |0,\sigma\ra &\rightarrow&  \cos(Jt) |0,\sigma\ra + i \sin(Jt) |\sigma,0\ra.
\end{eqnarray}
At a deterministic time $t_h=\frac{\pi}{2J}$, atom is completely
hopped to the next site leaving its initial site empty. When
tunneling is completed after $t=t_h$, by performing a phase shift
$\phi=\pi/2$ to the lattice potential, all the large (small)
barriers in the potential are replaced by small (large) ones, as
shown in Fig.~\ref{fig1}(b), and atom thus resides in a new double
well. So, the dynamics of Eq.~(\ref{single-site-evolution}) is
repeated and atom hops one site further. Repeating this process
for $N-1$ times transports the atom from site $1$ to site $N$ and
with choosing the initial phase $\phi$ one can determine the
direction of transport towards the right or left. According to
Eq.~(\ref{single-site-evolution}) at each hopping step, there is a
global phase which is independent of internal state $\sigma$ and
is neglected in the rest of the paper.
In the case that both atoms sit in even
or odd sites they hop together in the same direction and thus
never reach each other. To bring them together one can load up one of the atoms while after a hopping time another atom tunnel to its neighboring site. By this mechanism, two atoms sit in even-odd sites and can be transported towards each other.

 In this common double well atoms interact and
thus the quantum gate is realized. The Hilbert space of two atoms
in a double dot is spanned by $|\sigma,\sigma'\ra$,
$|0,\sigma\sigma'\ra$ and $|\sigma\sigma',0\ra$ where, 0
represents an empty site, $\sigma,\sigma'=\uparrow,\downarrow$ and
two sites are separated by ``,". The evolution can be written as
\begin{eqnarray}\label{double-psi-t}
  |\sigma,\sigma\ra \rightarrow &A_{\sigma,\sigma}(t)&|\sigma,\sigma\ra + C_{\sigma,\sigma}(t) (|\sigma\sigma,0\ra+|0,\sigma\sigma\ra) \cr
  |\sigma,\sigma'\ra \rightarrow  &A_{\sigma,\sigma'}(t)&|\sigma,\sigma'\ra+ B_{\sigma',\sigma}(t) |\sigma',\sigma\ra +\cr
  &C_{\sigma,\sigma'}(t)& (|\sigma\sigma',0\ra+|0,\sigma\sigma'\ra), \ \ \text{for} \ \ \sigma\neq \sigma'
\end{eqnarray}
where
\begin{eqnarray}\label{parameters}
  A_{\sigma,\sigma} (t) &=& \cos(sUt/2)+\frac{i}{s} \sin(sUt/2) \cr
  C_{\sigma,\sigma} (t) &=& i\frac{2\sqrt{2}J}{sU} \sin(sUt/2) \cr
  A_{\sigma,\sigma'}(t) &=& \frac{e^{iUt/4}}{2}+\frac{1}{2}(\cos(s'Ut/4)+\frac{i}{s'}\sin(s'Ut/4)) \cr
  B_{\sigma,\sigma'}(t) &=& -\frac{e^{iUt/4}}{2}+\frac{1}{2}(\cos(s'Ut/4)+\frac{i}{s'}\sin(s'Ut/4)) \cr
  C_{\sigma,\sigma'}(t) &=& \frac{i4J}{s'U}\sin(s'Ut/4),
\end{eqnarray}
for $s=\sqrt{1+16J^2/U^2}$ and $s'=\sqrt{1 + 64 J^2/U^2}$. This
dynamics may look very complicated but in the limit of $J\ll U$,
where the double occupancy is negligible and Hamiltonian is
effectively an XX spin Hamiltonian with superexchange coupling $J_{ex}=4J^2/U$ \cite{Lukin}, is simplified to
\begin{eqnarray}\label{gates}
  |\sigma,\sigma\ra  &\rightarrow& |\sigma,\sigma\ra \cr
  |\sigma,\sigma'\ra &\rightarrow& \cos(J_{ex}t)|\sigma,\sigma'\ra+i\sin(J_{ex}t)|\sigma',\sigma\ra.
\end{eqnarray}
This dynamics explains a conditional gate acting on two qubits and
in particular after interacting for $t_I=\frac{\pi}{2J_{ex}}$ it
generates a maximally entangled state from a fully separable
initial state $|++\ra$ where,
$|+\ra=(|\uparrow\ra+|\downarrow\ra)/\sqrt{2}$. It is important to
notice that $U$ cannot be arbitrarily large as the interaction
time $t_I$ increases by increasing $U$. When interaction is accomplished, both atoms are
directed back to their initial sites and by switching on the local
traps they are returned back to their original confinements.

\section{Time Scale}  \label{sec4}

The total time needed for bringing two atoms
together from a distance $N$, let them interact to get entangled
and then restore them back in their initial positions is
$t_T=(N-2)t_h+t_I$. By considering the values of $t_h$ and $t_I$
one reads $t_T=\frac{\pi(N-2)}{2J}+\frac{\pi U}{8J^2}$. This time
scale is much faster than those from spin chains which scale as
$\sim N U/4J^2$ \cite{bayat-gate,bloch-phase-shift}, as only one
step during the process is governed by superexchange coupling
$J_{ex}$ and the rest is controlled by $J$. To have some
estimations for $t_T$ we take the parameters of a very recent
experiment \cite{Bloch-AC-driving} where, $J/h\simeq 1.5$ KHz and
$J_{ex}/h \simeq 250$ Hz (these give $U\simeq 25J$). For a
distance of $N=100$, taking these values gives us $t_T\simeq 17$
ms which is well below the typical decoherence time of the
internal levels ($\simeq 10$ minutes \cite{bollinger}). For
comparison, we use the same parameters and take the proposal of
Ref.~\cite{bloch-phase-shift}, which uses superexchange
interaction to distribute entanglement, that gives the time scale
of $\sim 100$ ms for the distance of 100 sites.  This shows that
our mechanism is faster by at least a factor of 5.

\section{Imperfections} \label{sec5}

In any experiment there might be some imperfections for realizing
a theoretical idea. For the above proposal, these imperfections
introduce as follows.

\subsection{Double occupancy}
The first imperfection effect is double occupancy, where two atoms
occupy the same site, when they reach to a common double well. When the initial state of two atoms in a
double well is $|++\ra$ this probability becomes
$P_{doub}=(|C_{\uparrow\uparrow}|^2+|C_{\downarrow\downarrow}|^2+
|C_{\uparrow\downarrow}|^2+|C_{\downarrow\uparrow}|^2)/4$. A
typical value of $U=25J$ \cite{Bloch-AC-driving} which is used in all our simulation, gives $P_{doub}
< 0.02$ which guarantees that the double occupancy is negligible.
It is worthwhile to mention that during the
transport part of our mechanism there is no need for large $U$ as
we do not have two atoms in a single double well and only during
the interaction time lattice potential should be tuned for large $U$.

\subsection{Leakage}

In reality, it is very difficult to have a series of fully
decoupled double wells as any non-zero tunneling through the
higher barriers diffuses the wave function and spoils both the
localization of the wave function and the deterministic time
evolution of the transport. To see this effect we assume that the
even and odd couplings are collectively switched from
$J_{max}=J(1+\delta)/2$ to $J_{min}=J(1-\delta)/2$ and vice versa,
where, $\delta$ is a dimensionless parameter less than 1. The
ideal situation, explained in previous sections, corresponds to
$\delta=1$ which gives the desired values of $J_{max}=J$ and
$J_{min}=0$. Our numerical simulations, however, show that for
$\delta=0.99$ the probability of transferring a particle over a
distance of $N=100$ remains above $0.9$. The couplings of the
system are related to the potential barriers through \cite{Lukin}
\begin{eqnarray}\label{Jmax_Jmin}
    J_{min} &=& \frac{4}{\sqrt{\pi}} \{ E^R_l (V_s+V_l)^3 \} ^{1/4}
    e^{-2\sqrt{(V_s+V_l)/E^R_l}}, \cr
    J_{max} &=& \frac{4}{\sqrt{\pi}} (E^R_s V_s^3)^{1/4}
    e^{-2\sqrt{V_s/E^R_s}},
\end{eqnarray}
where, $E^R_j=h^2/2m\lambda_j^2$ (for $j=s,l$) are recoil energies
in which $m$ is the atomic mass.  For typical experimental values
of $V_l=40 E_l^R$ and $V_s=7E_s^R=28E_l^R$
\cite{Stefan_Trotzky,Cheinet,Bloch-AC-driving} we get
$J_{max}/J_{min}\simeq 200$ which corresponds to $\delta=0.99$ and
the efficiency remains over 0.9 even for a transport over a large
distance of $N=100$ sites.

\subsection{Excitation of higher bands}
\begin{figure} \centering
    \includegraphics[width=8cm,height=6cm,angle=0]{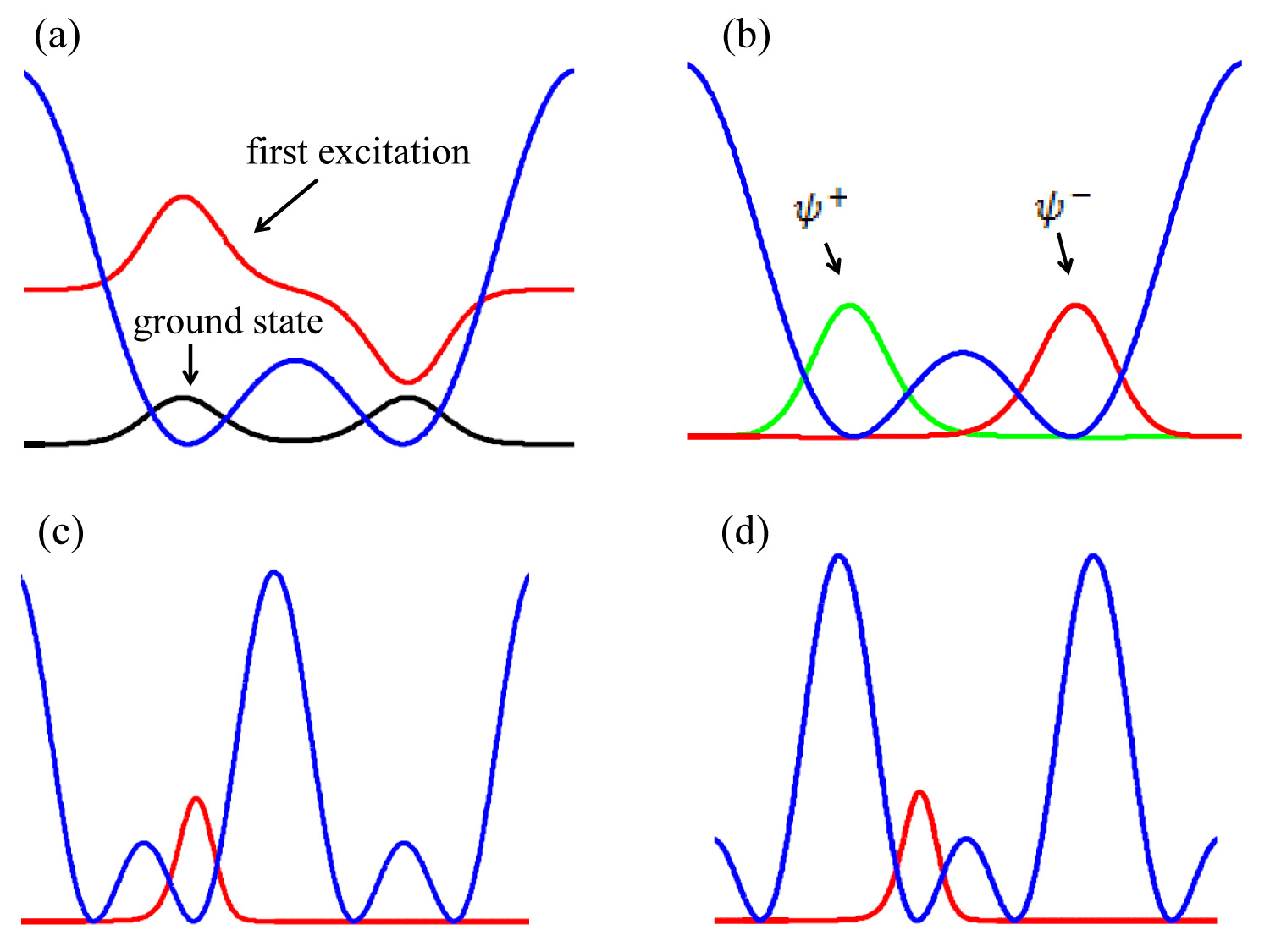}
    \caption{(Color online) (a) The wave function of the ground and the first excited state of a double well.
    (b) $\psi^{\pm}$ as the superposition of the ground and first excited wave functions.
    (c) The localized wave function before phase shift operation.
    (d) The ideal scenario in which the wave function remains unaffected after the phase shift operation.}
     \label{Fig3}
\end{figure}

\begin{figure} \centering
    \includegraphics[width=8cm,height=6cm,angle=0]{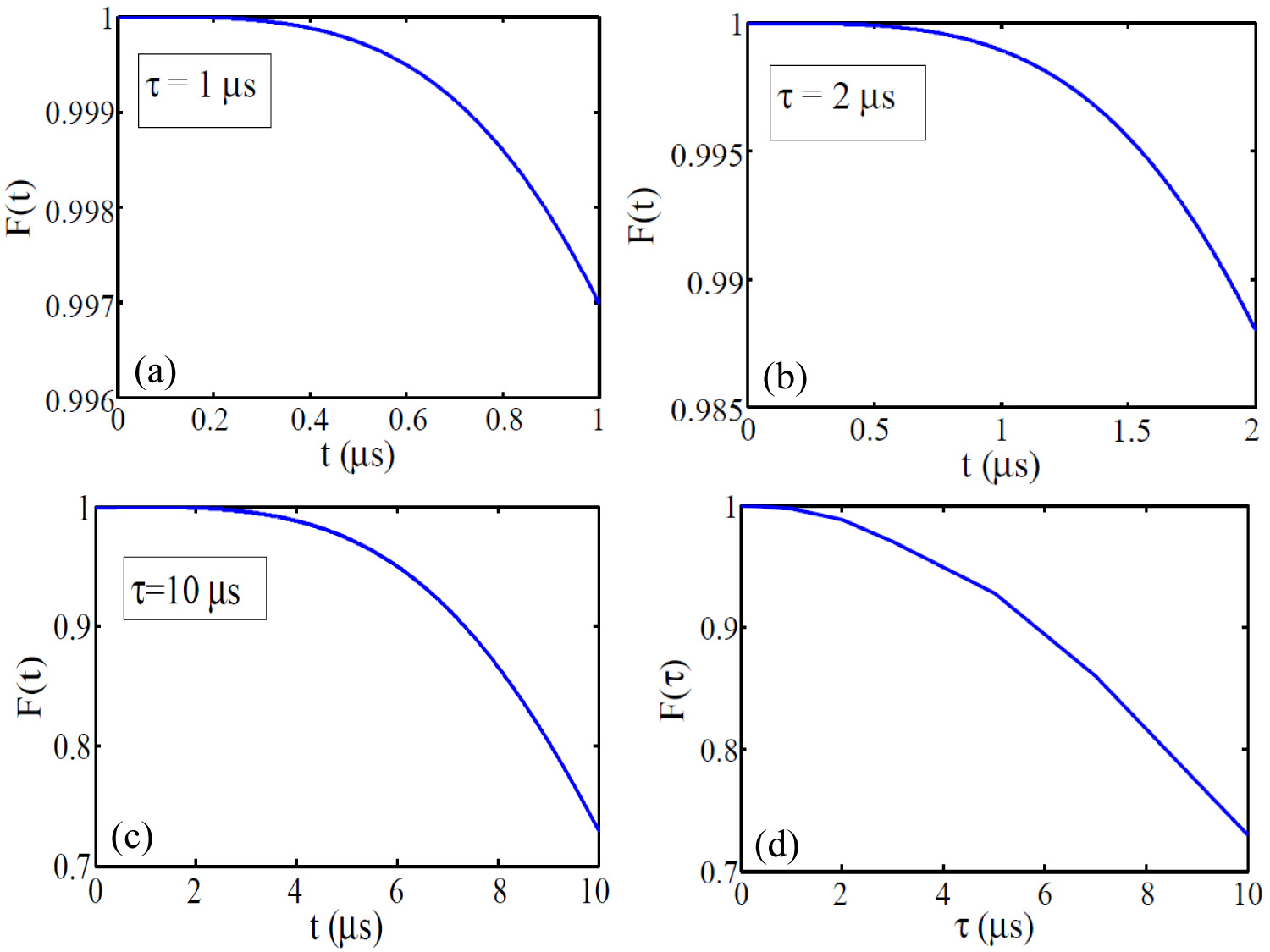}
    \caption{(Color online) The variation of fidelity $F$ vs. the time for (a) $\tau =1$ $\mu$s;
    (b) $\tau =2$ $\mu$s; and (c) $\tau =10$ $\mu$s. (d) The final fidelity $F(\tau)$ versus the switching time $\tau$ .}
     \label{Fig4}
\end{figure}

The most important imperfection in our mechanism is perhaps
exciting the higher energy vibrational modes during the operation
of phase shifts. If higher modes are excited then we get various
hopping rates for different vibrational eigenstates and thus the
transport of atoms cannot simply be determined by a single
coupling $J$. As the result, the wave function defuses to other
sites and the protocol fails to give deterministic transport of
localized atoms, the main objectives of the whole procedure. To
quantify the details of this effect, evolution of the localized
atomic wave function is considered during the operation of the
phase shift. The spatial part of the wave function for both the
ground and the first excited states of a double well, formed by typical
experimental values of $V_l=40 E_l^R$ and $V_s=7E_s^R$ \cite{Stefan_Trotzky,Cheinet,Bloch-AC-driving}, is plotted
in the Fig. \ref{Fig3}(a) where, $E_l^R$ and $E_s^R$ are the recoil energies. A localized wave function can be
considered as the superposition of these two low lying
eigenvectors as $|\psi ^{\pm}\rangle=\frac{1}{\sqrt{2}}(|g\rangle
\pm |e\rangle)$. The spatial parts of these states are plotted in
Fig. \ref{Fig3}(b). To see how system evolves during the
application of the phase shift we determine the evolution of the
wave function of a single atom by solving the time dependent
Schr\"{o}dinger equation
\begin{equation}\label{Schrodinger-Eq}
    i\hbar\frac{\partial \psi(x,t)}{\partial t} = -\frac{\hbar^2}{2m} \frac{d^2\psi(x,t)}{dx^2} + V(x,t)\psi(x,t),
\end{equation}
where $V(x,t)$ is given in Eq.~(\ref{V_lattice}) with $\phi$ linearly changes from 0 to $\pi/2$ over the time scale of $\tau$ and the
the initial condition is $\psi(x,0)=\psi^+(x)$. Ideally, during the phase shift process the wave function remains unaffected, as shown in Fig.~\ref{Fig3}(c) and (d),  and should have zero overlap with higher energy levels. To quantify the quality of the operation we calculate the
fidelity between $\psi(x,t)$ and $\psi(x,0)$ as
\begin{equation}\label{Fidelity}
    F(t)=|\int\psi^*(x,t)\psi(x,0) dx|.
\end{equation}
In Figs. \ref{Fig4}(a)-(c) we plot the fidelity $F(t)$ as a function of time within the time interval of $[0,\tau]$ for different values of $\tau$. According to
these results, the final fidelity, i.e. $F(\tau)$, decreases with increasing $\tau$. In Fig. \ref{Fig4}(d), the final fidelity $F(\tau)$ is plotted versus
$\tau$ which evidently shows that to have a fidelity larger than 0.95 the phase shift operation should be accomplished in less than 5 $\mu$s which seems fully accessible to current experiments \cite{Meschede}.

\begin{figure} \centering
    \includegraphics[width=8cm,height=3.5cm,angle=0]{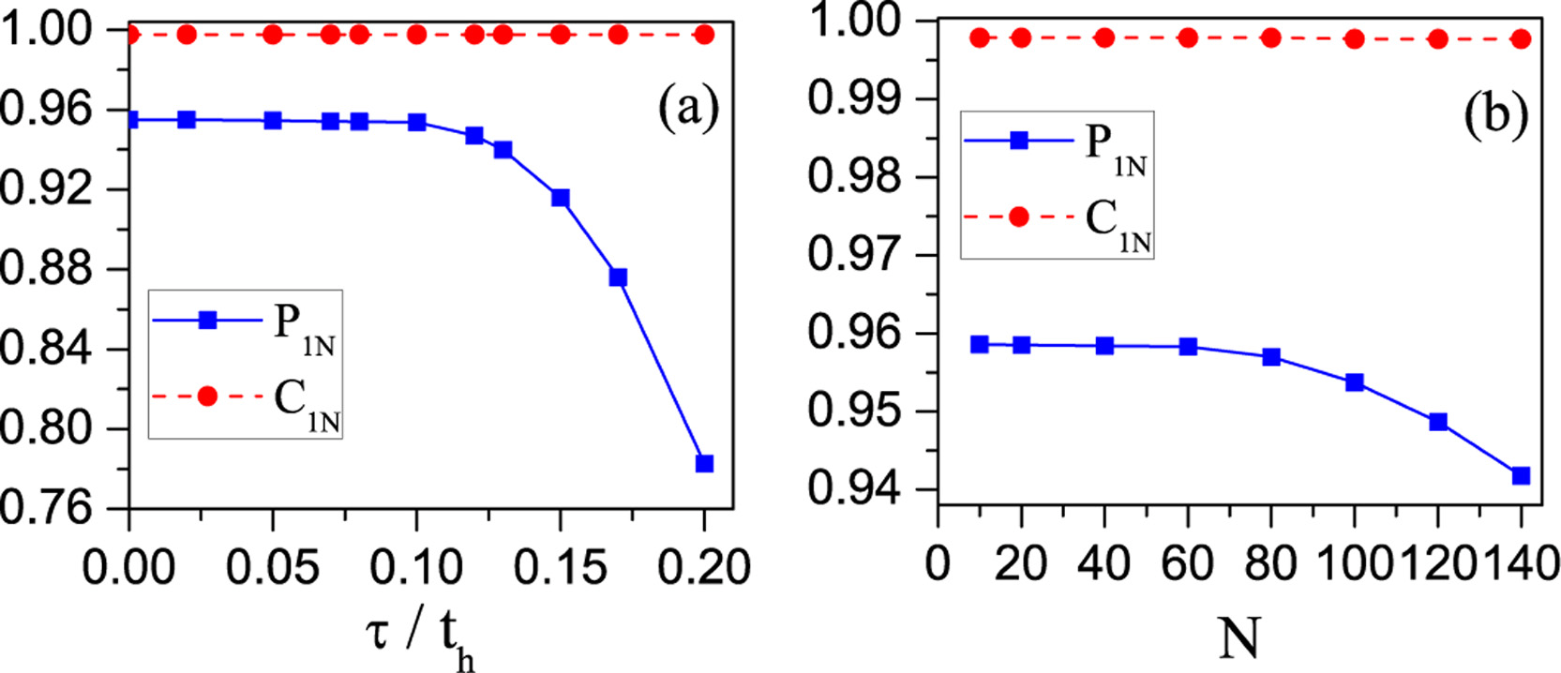}
    \caption{(Color online) (a) $P_{1N}$ and $C_{1N}$ versus switching time $\tau/t_h$ for a chain of length $N=100$. (b) $P_{1N}$ and $C_{1N}$ versus distance $N$ for the switching time of $\tau=0.1 t_h$ (in both figures $U=25J$).}
     \label{fig2}
\end{figure}

\subsection{Gradual switching }
So far, we have assumed that applying the phase shift $\phi$ is
instantaneous, however, in real experiments it
takes some time to perform this phase shift
\cite{Mandel-gate,Meschede-gate}. To see this effect we assume
that when odd (even) couplings, i.e. $J_0$ ($J_e$), change from
$J$ ($0$) to $0$ ($J$) this transition is linear in time and
happens over a time period of $\tau$. Similarly, when odd (even)
couplings change from $0$ ($J$) to $J$ ($0$) we assume a linear
rise (fall) over the same time period $\tau$. The gradual
switching of the couplings couples different double wells along
the lattice and delocalizes the the atomic wave functions which
has two main effects: (i) atomic wave functions defuse over
several sites and thus atoms may not reach their destination with
certainty; (ii) entanglement generation may be affected as atoms
may not both be localized in the same double well to interact. To
quantify these effects in the presence of gradual switching of the
couplings we assume that two atoms in the state $|++\ra$,
initially localized in sites $1$ and $N$, are directed towards
each other to interact and generate entanglement. At $t_T$ we
expect to have both atoms returned to their initial positions in a
maximally entangled state, however, due to the effect of gradual switching they may not be localized at sites 1 and $N$. So, we project the state of the system to these two sites regardless of their internal states. The probability of this projection is $P_{1N}$ and determines the
probability of finding atoms at sites 1 and $N$. This projection can be done
experimentally by taking a fluorescent picture of the lattice at
$t=t_T$ \cite{single-site,Weitenberg} in which with probability $P_{1N}$ atoms are found in sites $1$ and $N$.
During the loading and fluorescence imaging the dipole trap for determination of atomic
Positions is upper than for manipulation of the atoms.
Furthermore, the particular exposure time of the fluorescence imaging is chosen depending on the required precision for the determination of atomic positions \cite{karski_lett} and also the wavelength and so the diffraction limit is chosen to avoid any interaction with the internal state of the atoms \cite{Meschede}.
Nevertheless, projecting atoms in their initial sites does not guarantee that they are entangled as due to the delocalization they may not reside in the same double well with certainty during the interaction period to interact and get entangled.
To see this effect we take the density matrix of the atoms after projection on sites 1 and $N$ and compute their concurrence $C_{1N}$ as a
measure of entanglement \cite{concurrence}. In other words, $C_{1N}$ quantifies the entanglement between the atoms provided that they are found in sites $1$ and $N$.

To have an
estimation of $\tau$ needed for applying the phase shift we take
the results of Ref.~\cite{Meschede-gate} in which our desired
phase shift is performed over the time of $\tau\simeq 15$ $\mu$s.
For a typical value of the tunneling such as $J/h=1.5$ KHz
reported in \cite{Bloch-AC-driving} we get $\tau/t_h < 0.1$. In
Fig.~\ref{fig2}(a) we plot both $P_{1N}$ and $C_{1N}$ in terms of
$\tau/t_h$ for a chain of length $N=100$ where, we have $N-1$ {\em
gradual} switching of the couplings. As it is clear from the
figure, there is plateau up to $\tau\simeq 0.1 t_h$ and beyond
that even for a very pessimistic value of $\tau/t_h=0.15$ we have
$P_{1N}>0.9$. Entanglement is even more robust and remains almost
one for all values of $\tau$ which shows that whenever atoms
return to their initial positions, with probability $P_{1N}$, entanglement is almost perfectly
achieved. In Fig.~\ref{fig2}(b) we plot $P_{1N}$ and $C_{1N}$
versus the distance $N$ when $\tau=0.1 t_h$. This shows that
system has a very good scalability as $P_{1N}\simeq 0.94$ for a
distance of $N=140$.

\subsection{Imperfect phase shift}
\begin{figure} \centering
    \includegraphics[width=7cm,height=5.5cm,angle=0]{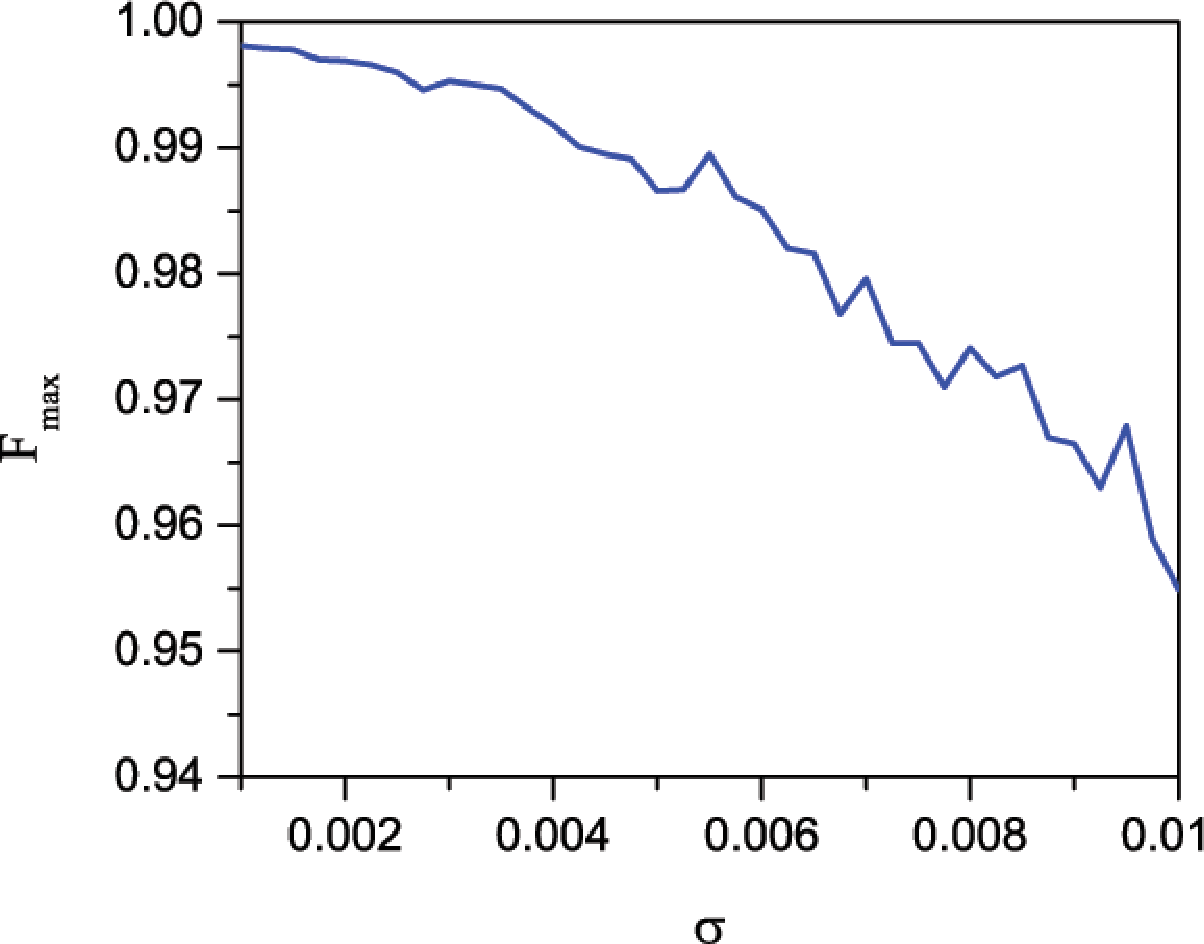}
    \caption{(Color online) Maximum amount of fidelity of 1 atom transfer from the site $1$ to site $N=100$ versus the different amounts of the random variable $\sigma$.}
     \label{phase-shift}
\end{figure}

 In reality it is difficult to apply phase shifts with exactly $\phi=0$ and $\phi=\pi/2$. As the result, at each phase shift operation there is an error in the couplings of the system such that the couplings change from $J(1-\epsilon)$  to $J\epsilon$ (or vice versa) in which $\epsilon$ is a dimensionless random number uniformly distributed between $0<\epsilon<\sigma$ and varies at each phase shift operation. To see the effect of this imperfect phase shift, the fidelity of a single atom transport from site $1$ to site $N=100$ has been calculated versus $\sigma$ by averaging over 100 different realizations at each value of $\sigma$. The result is plotted in Fig. \ref{phase-shift} in which the fidelity stays above $0.95$ for imperfections up to $\sigma=0.01$.

\subsection{Different raising times}
One may also worry about different raising times of the atoms from local traps and loading them into the optical lattice. This is not a big issue at all as the process of rasing the atoms and loading them to the optical lattice is an adiabatic process and should be accomplished very slowly \cite{Nakahara}. So, one can control the speed of this process in such a way that both atoms fully load to the optical lattice before starting their transmission. Furthermore, initially we can switch on a very deep lattice to avoid tunneling of the atoms and give them enough time to both reside in their position and then change the deep lattice to the desired superlattice given in Eq.~(\ref{V_lattice}) by changing the intensity and the phase of the laser beams.

\section{Comparison with AC-driven transport} \label{sec6}

In a periodic lattice, one may use AC-driving fields for breaking
certain temporal and spatial symmetries to get desymmetrized
Floquet states in the spectrum of the periodic Hamiltonian
\cite{Renzoni-PRL,Salgar-science,Denisov} and achieve an average rectified
current for an atomic cloud ($\sim 10^4$ atoms). For single atoms,
this does \emph{not} give a ``deterministic" transport in which
atoms remain ``localized", the two key features essential for the
proposed quantum computer. At variance, as discussed in the
previous section, realistic situation for keeping mobile atoms
localized during their fast deterministic transport.

One may also use high frequency Ac-driving fields for
\emph{renormalizing} the tunneling parameters
\cite{creffield-normalized} without breaking symmetries. With a
sinusoidal driving of amplitude $K$ and frequency $\omega$ the
renormalization factor is $\mathcal{J}_0(Kx/\omega)$ where,
$\mathcal{J}_0$ is the zeroth order Bessel function of type one
and $x$ is the distance between the two subsequent minima. In
Refs. \cite{creffield-gate,quantum-rachet}, this mechanism is used
for switching off the even or odd couplings alternatively to
achieve directed transport of atoms or quantum states. In this
approach different distances $x_1$ and $x_2$ between alternating
minima are used such that $\mathcal{J}_0(Kx_1/\omega)\neq 0$ when
$\mathcal{J}_0(Kx_2/\omega)= 0$ and vice versa. So, when one set
of couplings (even or odd) are switched off, i.e.
$\mathcal{J}_0(Kx_2/\omega)= 0$, the other set of couplings are
also renormalized by the small number
$\mathcal{J}_0(Kx_1/\omega)$. Consequently, this results in a slow
evolution which practically shows no benefit over spin chain
superexchange dynamics \cite{bloch-phase-shift,bayat-gate}.

\section{Entanglement Detection} \label{sec7}
An entanglement witness is defined as an Hermitian operator such that its
expectation value is positive for every separable state but negative for some
entangled states \cite{terhal} and will have
at least one negative eigenvalue \cite{terhal_2}. Given these properties, this observable $W$ reveals the presence of entanglement in mixture
$\rho^{AB}$ on the composite system space $\mathcal{H_A} \otimes \mathcal{H_B}$
such that $tr(\rho^{AB} W) < 0$ for entangled state but $tr(\rho^{AB}W)\geq 0$ for all separable states \cite{Horodecki}.
To detect entanglement between atoms residing in sites $1$ and
$N$, we introduce an entanglement witness which its average value
is negative for entangled states and positive otherwise. Following
the footsteps of the Ref.~\cite{horodecki-separability} gives us
\begin{equation}\label{witness}
    W= \frac{1}{2}\{I-Z_1Y_N- Y_1Z_N -X_1X_N\},
\end{equation}
where $X_k$, $Y_k$ and $Z_k$ are Pauli operators acting on qubit
$k$ and $I$ is identity operator.
Local unitary operations on individual atoms
\cite{single-site,Weitenberg} and single spin measurement
\cite{Meschede} are now available for atoms in optical lattices. However,
due to practical problems spin measurement might be restricted to
$Z$ measurements \cite{Meschede}. So, to have only $Z$ operation
in the witness $W$ we have to transform $X$ and $Y$ Pauli
operators by local rotations as $X = h Z h$ and
$Y=e^{i\frac{\pi}{4} X} Z  e^{-i\frac{\pi}{4} X}$ where,
$h=(X+Z)/\sqrt{2}$ is the Hadamard gate. Notice that these local
operations do not need tight laser beams as the optical lattice is
almost empty.

\section{Summary} \label{sec8}
We introduced a fully coherent, fast and easily realizable method
for directed transport of localized atoms in optical
superlattices. Our proposal is based on performing phase shifts
regularly during the free evolution of the system. The features of
our proposal can be summarized as: (i) Atoms remain localized
during their transport. Transporting process is also (ii) coherent; (iii) fast and (iv) easily realizable
in the laboratory. These features help to realize a possible
quantum computer using local optical traps and an empty optical
lattice. In contrast to superexchange interaction in spin chains
(time scale $\sim NU/J^2$), our coherent mechanism is fast and
works within the time scale of $\sim N/J$ and does not need
sophisticated cooling mechanisms. Furthermore, our proposal is
fundamentally different from the widely studied ratchet effect
\cite{Hanggi,Renzoni-PRL,Salgar-science,Denisov}, caused by breaking
certain temporal and spatial symmetries.  In fact, in our protocol
symmetries of the Hamiltonian are not broken and it is the initial
state which is desymmetrized. Unlike the usual ratchet proposals,
our mechanism is fully deterministic and applicable for localized
individual atoms. Our proposal is also distinct by its faster
speed in compare to the other AC-driven transports which, similar
to ours, do not break the symmetries and instead renormalize the
tunnelings of the itinerant atoms
\cite{creffield-gate,quantum-rachet}. In addition, we showed that
our proposed scenario is robust against several imperfections
which makes it accessible to current experiments and can be
realized immediately with current technology.

\section{Acknowledgments} \label{sec9}
Discussion with S.~Bose, C.~E.~Creffield,
J.~Denschlag, M.~B.~Plenio and A.~Retzker, are kindly
acknowledged. MR thanks M. Soltani and H. Mohammadi for their
hospitality in University of Isfahan. AB thanks the Alexander von
Humboldt Foundation and the EU STREPs CORNER, HIP and PICC for
their support.


\begin{thebibliography}{99}
\bibitem{Bremner} M. J. Bremner, Ch. M. Dawson, J. L. Dodd, A. Gilchrist, A. W. Harrow, D. Mortimer, M. A. Nielsen and T. J. Osborne, Phys. Rev. Lett. {\bf 89}, 247902 (2002).

\bibitem{albanese} C. Albanese, M. Christandl, N. Datta and A. Ekert, Phys. Rev. Lett. {\bf 93}, 230502 (2004);S. R. Clark, C. M. Alves, D. Jaksch, New J. Phys. {\bf 7}, 124 (2005); T.J. Osborne and N. Linden, Phys. Rev. A {\bf 69}, 052315 (2004).

\bibitem{Yung-bose} M.~H. Yung and S. Bose, Phys. Rev. A {\bf 71}, 032310 (2005);
M.~H. Yung, S. C. Benjamin and S. Bose, Phys. Rev. Lett. {\bf 96},
220501 (2006).

\bibitem{bayat-gate} L.~Banchi, A.~Bayat, P.~Verrucchi, and S.~Bose, Rev. Lett. {\bf 106}, 140501 (2011).

\bibitem{Lukin-gate} N. Y. Yao, L. Jiang, A. V. Gorshkov, Z.-X. Gong, A. Zhai, L.-M. Duan and M. D. Lukin, Phys. Rev. Lett. {\bf 106}, 040505 (2011);
N. Y. Yao, L. Jiang, A. V. Gorshkov, P. C. Maurer, G. Giedke, J. I. Cirac and M. D. Lukin, Nat. Commun. {\bf 3}, 800 (2012).

\bibitem{WojcikLKGGB} A. W{\'o}jcik, T. Luczak, P. Kurzy\'{n}ski, A. Grudka, T. Gdala and M. Bednarska, Phys. Rev. A {\bf 72}, 034303 (2005).


\bibitem{Cirac-bus} J.~Cirac and P.~Zoller, Nature {\bf 404}, 579 (2000).

\bibitem{Hanggi} P. Reimann, Phys. Rep. {\bf 361}, 57 (2002); P. H\"{a}nggi and F.~Marchesoni, Rev. Mod. Phys. {\bf 81}, 387 (2009).

\bibitem{Renzoni-PRL} P. H. Jones, M. Goonasekera, and F. Renzoni, Phys. Rev. Lett. {\bf 93},
073904 (2004); R. Gommers, P. Douglas, S. Bergamini, M. Goonasekera, P. H. Jones, and F. Renzoni, Phys. Rev. Lett. {\bf 94}, 143001 (2005). S. Denisov, S. Kohler, P. H\"{a}nggi, EPL {\bf 85}, 40003 (2009).

\bibitem{Salgar-science} T. Salger, S. Kling, T. Hecking, C. Geckeler, L. M. Molina and M. Weitz, Science {\bf 326}, 1241 (2009);

\bibitem{Denisov} S. Denisov, L. M. Molina, S. Flach, and P. H\"{a}nggi, Phys. Rev. A {\bf 75}, 063424 (2007).

\bibitem{creffield-normalized} F. Grossmann, T. Dittrich, P. Jung, and P. H\"{a}nggi , Phys. Rev. Lett. {\bf 67}, 516 (1991);
M.~Holthaus, Phys. Rev. Lett. {\bf 69}, 351 (1992); F. Gro\ss mann and P. H\"{a}nggi, EPL {\bf 18}, 571 (1992); C. E. Creffield, Phys. Rev. B {\bf 67}, 165301 (2003); D. Zueco, F. Galve, S. Kohler and P. H\"{a}nggi, Phys. Rev. A {\bf 80}, 042303 (2009).


\bibitem{Bloch-AC-driving} Y.~A.~Chen, S. Nascimb\`{e}ne, M. Aidelsburger, M. Atala, S. Trotzky and I. Bloch, Phys. Rev. Lett. {\bf 107}, 210405 (2011).

\bibitem{creffield-gate} C.~E.~Creffield, Phys. Rev. Lett. {\bf 99}, 110501 (2007).

\bibitem{Isart} O.~R.~Isart, J.~J.~G.~Ripoll, Phys. Rev. A {\bf 76}, 052304 (2007).
\bibitem{quantum-rachet} K.~Hai, W.~Hai, and Q.~Chen, Phys. Rev. A {\bf 82}, 053412 (2010).


\bibitem{Calarco-rachet} G.~De~Chiara, T. Calarco, M. Anderlini, S. Montangero, P. J. Lee, B. L. Brown, W. D. Phillips and J. V. Porto, Phys. Rev. A {\bf 77}, 052333 (2008).

\bibitem{Bloch-Review} I. Bloch, Nature {\bf 453}, 1016 (2008).

\bibitem{Delgado} J. J. G. Ripoll, M. A. M. Delgado and J. I. Cirac, Phys. Rev. Lett. {\bf 93}, 250405 (2004).

\bibitem{Mott-insulator} W. S. Bakr, J. I. Gillen, A. Peng, S. F\"{o}lling, and M. Greiner, Nature {\bf 462}, 74 (2009);
M. Greiner, O. Mandel, T. Esslinger, T. W. H\"{a}nsch, and I. Bloch, Nature {\bf 415}, 39 (2002); W. S. Bakr,
A. Peng, M. E. Tai, R. Ma, J. Simon, J. I. Gillen, S. F\"{o}lling, L. Pollet, M. Greiner, Science {\bf 329}, 547 (2010); R. J\"{o}rdens, N. Strohmaier, K. G\"{u}nter, H. Moritz and T. Esslinger, Nature {\bf 455}, 204 (2008); U. Schneider, L. Hackerm\"{u}ller, S. Will, Th. Best, I. Bloch, T. A. Costi, R. W. Helmes, D. Rasch and A. Rosch, Science {\bf 322}, 1520 (2008).

\bibitem{Lukin} L. Duan, E. Demler and M. D. Lukin, Phys. Rev. Lett. {\bf 91}, 090402 (2003).

\bibitem{bloch-phase-shift} P.~Barmettler, A. M. Rey, E. Demler, M. D. Lukin, I. Bloch and V. Gritsev, Phys. Rev. A {\bf 78}, 012330 (2008).

\bibitem{Medley2010} P. Medley, D. M. Weld, H. Miyake, D. E. Pritchard and W. Ketterle, Phys. Rev. Lett. {\bf 106}, 195301 (2011).

\bibitem{greiner-AFM}  J.~Simon, W. S. Bakr, R. Ma, M. E. Tai, P. M. Preiss and M. Greiner, Nature {\bf 472}, 307 (2011).

\bibitem{single-site} J. F. Sherson, C. Weitenberg, M. Endres,	M. Cheneau, I. Bloch and S. Kuhr, Nature {\bf 467}, 68 (2010).

\bibitem{Weitenberg} C. Weitenberg, M. Endres, J. F. Sherson, M. Cheneau, P. Schau\ss, T. Fukuhara, I. Bloch and S. Kuhr, Nature {\bf 471}, 319 (2011).

\bibitem{karski_lett} M. Karski, L. F\"{o}rster, J. M. Choi, W. Alt, A. Widera, and D. Meschede, Phys. Rev. Lett. {\bf 102}, 053001 (2009).

\bibitem{Meschede} M. Karski, L. F\"{o}rster, J. -M. Choi, A. Steffen, N. Belmechri, W. Alt, D. Meschede and A. Widera, New J. Phys. {\bf 12}, 065027 (2010).

\bibitem{Gibbons-Nondestructive} M. J. Gibbons, C. D. Hamley, C. Y. Shih and M. S. Chapman, Phys. Rev. Lett. {\bf 106}, 133002 (2011). .

\bibitem{Jaeyoon-single-site} J. Cho, Phys. Rev. Lett. {\bf 99}, 020502 (2007).


\bibitem{Mandel-gate} O. Mandel, M. Greiner, A. Widera, T. Rom, Th. W. H\"{a}nsch and I. Bloch, Nature {\bf 425}, 937 (2003).

\bibitem{Meschede-gate} M. Karski, L. F\"{o}rster, J. -M. Choi, A. Alberti, W. Alt, A. Widera, D. Meschede, J. Korean Phys. Soc. {\bf 59}, 2947 (2011).

\bibitem{vaucher} B. Vaucher, S. R. Clark, U. Dorner and D. Jaksch New J. Phys {\bf 9}, 221 (2007); B. Vaucher, A. Nunnenkamp and D. Jaksch, New J. Phys. {\bf 10}, 023005 (2008).

\bibitem{Sebby-Strabley} J. Sebby-Strabley, M. Anderlini, P. S. Jessen, and J. V. Porto, Phys. Rev. A {\bf 73}, 033605 (2006).


\bibitem{Nakahara} M.~Nakahara, T.~Ohmi and Y.~Kondo, arXiv:1009.4426; E.~H.~Lapasar, K.~Kasamatsu, Y.~Kondo, M.~Nakahara, T.~Ohmi, J. Phys. Soc. Jpn. {\bf 80}, 114003 (2011).

\bibitem{terhal} M. B. Terhal, Theor. Comp. Sci. {\bf 287}, 313 (2002).

\bibitem{terhal_2} M. B. Terhal, Lin. Algebr. Appl. {\bf 323}, 61 (2001).

\bibitem{Horodecki} P. Horodecki, Phys. Lett. A {\bf 232}, 333 (1997).

\bibitem{horodecki-separability} M. Horodecki, P. Horodecki and R. Horodecki. Phys. Lett. A {\bf 223}, 1 (1996).

\bibitem{bollinger} J. J. Bollinger, D. J. Heizen, W. M. Itano, S. L. Gilbert and D. J. Wineland, IEEE Trans. on Instrum. and Measurement {\bf 40}, 126 (1991).

\bibitem{concurrence} W. K. Wootters, Phys. Rev. Lett., {\bf 80}, 2245 (1998).

\bibitem{Cheinet}P. Cheinet, S. Trotzky, M. Feld, U. Schnorrberger, M. Moreno-Cardoner, S. F\"{o}lling, and I. Bloch, Phys. Rev. Lett. {\bf 101}, 090404 (2008).

\bibitem{Stefan_Trotzky}S. Trotzky, Y. -A. Chen, U. Schnorrberger, P. Cheinet, and I. Bloch, Phys. Rev. Lett. {\bf 105}, 265303 (2010).


\end{thebibliography}
\end{document}